\documentclass[preprint,11pt]{article}
\usepackage{amssymb}
\title{Graviton Excitations and Lorentz-Violating Gravity with Cosmological Constant}

\author{\textbf{J.L. Boldo$^{*}$, J.A. Helay\"{e}l-Neto $^{*}$ L.M. de Moraes$^{*}$ ,}\\ \textbf{C.A.G. Sasaki$^{\dagger }\,$ and V.J. V\'{a}squez Otoya$^{\ddagger}$},\\
\\
$^{\ast}$CBPF, Centro Brasileiro{\bf \ }de Pesquisas F\'{\i}sicas \\
Rua Xavier Sigaud 150, 22290-180, Urca \\
Rio de Janeiro, Brazil\\
\\
$^{\dagger }$ UERJ, Universidade Estadual do Rio de Janeiro,\\
Departamento de Estruturas Matem\'{a}ticas\\
Instituto de Matem\'{a}tica e Estat\'{\i }stica\\
Rua S\~{a}o Francisco Xavier, 524 \\
20550-013, Maracan\~{a}, Rio de Janeiro, Brazil\\
\\
$^{\ddagger}$UFF, Universidade Federal Fluminense \\
Campus da Praia Vermelha, Gragoata, 24210-310, Niteroi \\
Rio de Janeiro, Brazil}
\begin{document}
\maketitle
\begin{abstract}
Motivated by the interest raised by the problem of Lorenz-symmetry violating gauge theories in connetion with gravity models, this contribution sets out to provide a general method to systematically study the excitation spectrum of gravity actions which include a Lorentz-symmetry breaking Chern-Simons-type action term for the spin connection. A complete set of spin-type operators is found which accounts for the (Lorentz) violation parameter to all orders and graviton propagators are worked out in a number of different situations.
\end{abstract}


\section{Introduction}

Theories with Lorentz symmetry breaking turned into a highly interesting research activity over the latest years. This may be motivated by the observation that  superstring theories suggest that,  at a certain energy scale, Lorentz symmetry should be violated \cite{Kostelecky:1988zi}; properties of these models such as CPT-symmetry violation and  vacuum birefringence \cite{Carroll:1989vb} could be related to some cosmological effects like cosmic magnetism \cite{Bertolami:1998dn,Campanelli:2008xs} and anisotropy of the CMB \cite{Kostelecky:2008be}
. \\ 
On the other hand, gravity theories with Lorentz symmetry breaking had been mainly exploited in \cite{Bluhm:2007bd,Jackiw:2003pm,Berezhiani:2007zf,Kostelecky:2009zr}; in contrast with the electromagnetic case, there is no vacuum birefingence and the diffeomorfism invariance is broken, so that the vacuum structure for graviton excitations is non-trivial and a production of massive gravity modes from the violation of Lorentz symmetry may become an issue of relevance. \\

At this point, it is valid to ask which kind of spectrum this new vacuum structure offers us. With this motivation, we provide a general method to compute the propagator for  gravity theories with Lorentz symmetry breaking based on the irreducible group decomposition  and the spin operator formalism developed by Barnes-Rivers \cite{BR} and Sezgin and van Nieuwenhuizen \cite{Sezgin:1979zf}. Thus, in agreement with this method, we list a acomplete set of spin projection-type operators which form a closed algebra; the wave operator is expanded in terms of this basis and the calculational procedure to read off the propagators becomes very systematic. Special cases of interest are finally contemplated. We do not consider possible effects of torsion: we assume that all gravity degrees of freedom are accomodated into the metric fluctuations.\\

This work is organized as follows: In Section 2, we introduce our notations, conventions and  start with a general action including the Einstein-Hilbert term, a second-order curvature term and the cosmological constant; then, we implement the weak field aproximation to carry out the analysis of the gravity excitations.  In Section 3, we proceed to calculate the propagators. Finally, in Section 4, we close with some Concluding Comments about the particle espectrum in some special cases. 

\section{Preliminaries}

We propose to carry out our analysis by starting off from the following
action for topologically massive gravity in four dimensions: 
\begin{equation}
S=\int d^{4}x\left[ \frac{\sqrt{-g}}{\kappa ^{2}}\left( -{\cal R}+\widetilde\Lambda +%
\frac{\sigma }{2}R^{2}+\frac{\xi }{2}R_{\mu \nu
}R^{\mu \nu }\right) {\cal +L}_{cs}\right] ,  \label{1}
\end{equation}
where 
\begin{equation}
{\cal L}_{cs}=-\frac{1}{2}\varepsilon ^{\mu \nu \kappa \lambda }v_{\mu
}\Gamma _{\lambda \sigma }^{\,\,\,\,\,\,\,\,\rho }(\partial _{\nu }\Gamma
_{\rho \kappa }^{\,\,\,\,\,\,\sigma }+\frac{2}{3}\Gamma _{\nu \alpha
}^{\,\,\,\,\,\,\,\sigma }\Gamma _{\kappa \rho }^{\,\,\,\,\,\,\alpha })
\label{2}
\end{equation}
is the topological Chern-Simons term,  and $\widetilde\Lambda=\kappa^2\Lambda$, $\Lambda$ being the cosmological constant and $\sigma$ and $\kappa$  coupling constants of $m^{-2}$ dimension. The idea of writing down the cosmological constant as above is simply for the sake of factoring out a $\kappa^{-2}$-factor, which simplifies the task of writing down the expressions for the propagators, to be calculated in the next Section.

We adopt here the Minkowski metric $\eta _{\mu \nu }=(+;-,-,-)$ and the
Ricci tensor, ${\cal R}_{\mu \nu }={\cal R}_{\lambda \mu \nu
}^{\,\,\,\,\,\,\,\,\,\,\,\,\lambda }$. In the Riemannian space-time, the coefficients of the affine connection are
expressed in terms of the Christoffel symbols, ($\left\{ _{\mu \nu
}^{\,\,\lambda }\right\} $), which are completely determined by the metric: 
\begin{equation}
\Gamma _{\mu \nu }^{\,\,\,\,\,\,\,\lambda }=\left\{ _{\mu \nu }^{\,\,\lambda
}\right\} =\frac{1}{2}g^{\kappa \lambda }\left( \partial _{\mu }g_{\kappa
\nu }+\partial _{\nu }g_{\mu \kappa }-\partial _{\kappa }g_{\mu \nu }\right).
\label{4}
\end{equation}
In order to derive the propagators and, consequently, the particle
spectrum of the theory, we linearize the metric-dependent part of the
Lagrangian by adopting the usual splitting in the weak gravitational field approximation: 
\begin{equation}
g_{\mu \nu }(x)=\eta _{\mu \nu }+\kappa h_{\mu \nu }(x),  \label{8}
\end{equation}
where $\kappa h_{\mu \nu }$ represents, as usually, the small pertubation around flat Minkowski
space-time.

The action is invariant under general coordinate transformations, 
\begin{equation}
\delta h_{\mu \nu }(x)=\partial _{\mu }\xi _{\nu }(x)+\partial _{\nu }\xi
_{\mu }(x),  \label{9}
\end{equation}
where $\xi _{\mu }$ is the gauge parameter. Therefore, it is necessary to
fix this gauge invariance in order to make the wave operator of the
Lagrangian non-singular. This is done by adding the usual De Donder
gauge-fixing term, 
\begin{equation}
{\cal L}_{gf}=\frac{1}{2\alpha }F_{\mu }F^{\mu },  \label{10}
\end{equation}
with 
\begin{equation}
F_{\mu }=\partial _{\nu }\left( h_{\,\,\,\,\mu }^{\nu }-\frac{1}{2}\delta
_{\,\,\,\,\mu }^{\nu }h\right) ,  \label{11}
\end{equation}
and $h\equiv h_{\mu }^{\,\,\,\mu }$. In this case, the action is the sum of
Einstein, Chern-Simons and gauge-fixing terms. So, by making use of the weak
field approximation for the metric, the bilinear terms can be collected as
below: 
\begin{eqnarray}\label{eq:12}
\left. {\cal L}\right.  &=&-\frac{1}{2}\left( \frac{1}{2}h^{\mu \nu }\Box
h_{\mu \nu }-\frac{1}{2}h\Box h+h\partial _{\mu }\partial _{\nu }h^{\mu \nu
}-h^{\mu \nu }\partial _{\mu }\partial _{\lambda }h_{\,\,\,\,\,\nu
}^{\lambda }\right)  \\
&&+\frac{\sigma }{2}\left( h\Box ^{2}h-2h\Box \partial _{\mu }\partial _{\nu
}h^{\mu \nu }+h^{\mu \nu }\partial _{\mu }\partial _{v}\partial _{\kappa
}\partial _{\lambda }h^{\kappa \lambda }\right)   \nonumber \\
&&+\frac{\xi }{8}\left( h^{\mu \nu }\Box ^{2}h_{\mu \nu }+h\Box ^{2}h-2h\Box
\partial _{\mu }\partial _{\nu }h^{\mu \nu }+2h^{\mu \nu }\partial _{\mu
}\partial _{v}\partial _{\kappa }\partial _{\lambda }h^{\kappa \lambda
}\right)   \nonumber \\
&&+\frac{\Lambda \kappa ^{2}}{4}\left( -h^{\mu \nu }h_{\mu \nu }+\frac{1}{2}%
h^{2}\right)   \nonumber \\
&&+\frac{1}{2\alpha }\left[ -h^{\mu \nu }\partial _{\mu }\partial _{\lambda
}h_{\,\,\,\,\nu }^{\lambda }+h^{\mu \nu }\partial _{\mu }\partial _{\nu }h-%
\frac{1}{4}h\Box h\right]   \nonumber \\
&&-\frac{\kappa ^{2}}{4}\left[ \varepsilon ^{\mu \nu \kappa \lambda }v_{\mu
}\left( h_{\lambda }^{\,\,\,\rho }\Box \partial _{\nu }h_{\rho \kappa
}-h_{\lambda }^{\,\,\,\rho }\partial _{\nu }\partial _{\rho }\partial
_{\sigma }h_{\,\,\,\,\kappa }^{\sigma }\right) \right] .  \nonumber
\end{eqnarray}
Once the $h_{\mu\nu}$-propagator is calculated, the parameters $\sigma,\xi$ and the background vector, $v^\mu$, shall be suitably chosen in order to avoid ghosts or tachyons in the spectrum.  The structure of the propagator poles and their corresponding residues will indicate if there are non-physical modes induced by the higher derivative terms and the Lorentz-violating term.  
\section{Spin Operators and Graviton Propagator}

We now rewrite the linearized Lagrangian (\ref{eq:12}) in a more convenient
form, namely 
\begin{equation}
{\cal L}=\frac{1}{2}h_{\mu \nu }{\cal O}^{\mu \nu \kappa \lambda }h_{\kappa
\lambda },  \label{13}
\end{equation}
where ${\cal O}_{\alpha \beta }$ is the wave operator. The propagator is
given by 
\begin{equation}
\left\langle 0\right| T\left[ h_{\mu \nu }\left( x\right) h_{\kappa \lambda
}\left( y\right) \right] \left| 0\right\rangle =i\left( {\cal O}^{-1}\right)
_{\mu \nu \kappa \lambda }\delta ^{4}\left( x-y\right) .  \label{14}
\end{equation}

In order to invert the wave operator, we shall use an extension of the
spin-projection operator formalism, where one needs to add now other new operators coming from the Chern-Simons
term. The operators for rank-2 symmetric tensors are given by: 
\begin{eqnarray}
P_{\mu \nu ,\kappa \lambda }^{(2)} &=&\frac{1}{2}\left( \theta _{\mu \kappa
}\theta _{\nu \lambda }+\theta _{\mu \lambda }\theta _{\nu \kappa }\right) -%
\frac{1}{3}\theta _{\mu \nu }\theta _{\kappa \lambda },  \label{15} \\
P_{\mu \nu ,\kappa \lambda }^{(1)} &=&\frac{1}{2}\left( \theta _{\mu \kappa
}\omega _{\nu \lambda }+\theta _{\mu \lambda }\omega _{\nu \kappa }+\theta
_{\nu \kappa }\omega _{\mu \lambda }+\theta _{\nu \lambda }\omega _{\mu
\kappa }\right) ,  \nonumber \\
P_{\mu \nu ,\kappa \lambda }^{(0-s)} &=&\frac{1}{3}\theta _{\mu \nu }\theta
_{\kappa \lambda },\,\,\,\,\,\,\,\,\,P_{\mu \nu ,\kappa \lambda
}^{(0-w)}=\omega _{\mu \nu }\omega _{\kappa \lambda },  \nonumber \\
P_{\mu \nu ,\kappa \lambda }^{(0-sw)} &=&\frac{1}{\sqrt{3}}\theta _{\mu \nu
}\omega _{\kappa \lambda },\,\,\,\,\,\,\,\,P_{\mu \nu ,\kappa \lambda
}^{(0-ws)}=\frac{1}{\sqrt{3}}\omega _{\mu \nu }\theta _{\kappa \lambda }, 
\nonumber
\end{eqnarray}
with $\theta_{\mu\nu}$ and $\omega_{\mu\nu}$ the transverse and longitudinal projection operators for vectors, respectively.
Thus, the wave operator can be expanded in terms of the spin projection
operators, as it follows below: 
\begin{eqnarray}
{\cal O}_{\mu \nu ,\kappa \lambda } &=&\left( \frac{\xi \Box ^{2}}{4}-\frac{\Box }{2}-\frac{\widetilde\Lambda}{2}\right) P_{\mu \nu ,\kappa \lambda
}^{(2)}-\left( \frac{\Box }{2\alpha }+\frac{\widetilde\Lambda}{2}\right)
P_{\mu \nu ,\kappa \lambda }^{(1)}  \label{19} \\
&&+\left[ \left( 3\sigma +\xi \right) \Box ^{2}+\frac{(4\alpha -3)\Box }{%
4\alpha }+\frac{\widetilde\Lambda}{4}\right] P_{\mu \nu ,\kappa \lambda
}^{(0-s)}  \nonumber \\
&&+\frac{\sqrt{3}}{4}\left( \frac{\Box }{\alpha }+\widetilde\Lambda \right)
(P_{\mu \nu ,\kappa \lambda }^{(0-s\omega )}+P_{\mu \nu ,\kappa \lambda
}^{(0-\omega s)})-\frac{1}{4}\left( \frac{\Box }{\alpha }+\widetilde\Lambda \right) P_{\mu \nu ,\kappa \lambda }^{(0-\omega )}+\frac{\kappa ^{2}\Box 
}{4}S_{\mu \nu ,\kappa \lambda }.  \nonumber
\end{eqnarray}
Our analysis of the spin operators induced by the Lorentz-symmetry violating term yields the whole set of structures to be listed below:
\begin{eqnarray}
\Sigma_{\mu \nu }&=&v_{\mu }\partial _{\nu },\\
 \Lambda _{\mu \nu }&=&v_{\mu }v_{\nu },\\
S_{\mu \nu }&=&\varepsilon _{\mu \nu \kappa \lambda }v^{\kappa }\partial^{\lambda };
\end{eqnarray}
\begin{eqnarray}
S_{\mu \nu ,\kappa \lambda } &=&\frac{1}{2}\left( \theta _{\mu \kappa
}S_{\nu \lambda }+\theta _{\mu \lambda }S_{\nu \kappa }+\theta _{\nu \kappa
}S_{\mu \lambda }+\theta _{\nu \lambda }S_{\mu \kappa }\right) , \\
\Pi _{\mu \nu ,\kappa \lambda }^{(1-a)} &=&\frac{1}{2}\left( \theta _{\mu
\kappa }\Sigma _{\nu \lambda }+\theta _{\mu \lambda }\Sigma _{\nu \kappa
}+\theta _{\nu \kappa }\Sigma _{\mu \lambda }+\theta _{\nu \lambda }\Sigma
_{\mu \kappa }\right) , \\
\Pi _{\mu \nu ,\kappa \lambda }^{(1-b)} &=&\frac{1}{2}\left( \theta _{\mu
\kappa }\Sigma _{\lambda \nu }+\theta _{\mu \lambda }\Sigma _{\kappa \nu
}+\theta _{\nu \kappa }\Sigma _{\lambda \mu }+\theta _{\nu \lambda }\Sigma
_{\kappa \mu }\right) , \\
\Pi _{\mu \nu ,\kappa \lambda }^{(2)} &=&\frac{1}{2}\left( \theta _{\mu
\kappa }\Lambda _{\nu \lambda }+\theta _{\mu \lambda }\Lambda _{\nu \kappa
}+\theta _{\nu \kappa }\Lambda _{\mu \lambda }+\theta _{\nu \lambda }\Lambda
_{\mu \kappa }\right) ,
\end{eqnarray}
\begin{eqnarray}
\Pi _{\mu \nu ,\kappa \lambda }^{(SL)} &=&\frac{1}{2}\left( S_{\mu \kappa
}\Lambda _{\nu \lambda }+S_{\mu \lambda }\Lambda _{\nu \kappa }+S_{\nu
\kappa }\Lambda _{\mu \lambda }+S_{\nu \lambda }\Lambda _{\mu \kappa
}\right) , \\
\Pi _{\mu \nu ,\kappa \lambda }^{(S\Sigma -a)} &=&\frac{1}{2}\left( S_{\mu
\kappa }\Sigma _{\nu \lambda }+S_{\mu \lambda }\Sigma _{\nu \kappa }+S_{\nu
\kappa }\Sigma _{\mu \lambda }+S_{\nu \lambda }\Sigma _{\mu \kappa }\right) ,
\\
\Pi _{\mu \nu ,\kappa \lambda }^{(S\Sigma -b)} &=&\frac{1}{2}\left( S_{\mu
\kappa }\Sigma _{\lambda \nu }+S_{\mu \lambda }\Sigma _{\kappa \nu }+S_{\nu
\kappa }\Sigma _{\lambda \mu }+S_{\nu \lambda }\Sigma _{\kappa \mu }\right) ,
\\
\Pi _{\mu \nu ,\kappa \lambda }^{(S\omega )} &=&\frac{1}{2}\left( S_{\mu
\kappa }\omega _{\nu \lambda }+S_{\mu \lambda }\omega _{\nu \kappa }+S_{\nu
\kappa }\omega _{\mu \lambda }+S_{\nu \lambda }\omega _{\mu \kappa }\right) ,
\end{eqnarray}
\begin{eqnarray}
\Pi _{\mu \nu ,\kappa \lambda }^{(\theta \Sigma -a)} &=&\frac{1}{\sqrt{3}}%
\theta _{\mu \nu }\Sigma _{\kappa \lambda },\,\,\,\Pi _{\mu \nu ,\kappa
\lambda }^{(\theta \Sigma -b)}=\frac{1}{\sqrt{3}}\theta _{\mu \nu }\Sigma
_{\lambda \kappa }, \\
\Pi _{\mu \nu ,\kappa \lambda }^{(\Sigma \theta )} &=&\frac{1}{\sqrt{3}}%
\left( \Sigma _{\mu \nu }\theta _{\kappa \lambda }+\Sigma _{\nu \mu }\theta
_{\kappa \lambda }\right) , \\
\Pi _{\mu \nu ,\kappa \lambda }^{(\theta L)} &=&\frac{1}{\sqrt{3}}\theta
_{\mu \nu }\Lambda _{\kappa \lambda },\,\,\,\,\,\Pi _{\mu \nu ,\kappa
\lambda }^{(L\theta )}=\frac{1}{\sqrt{3}}\Lambda _{\mu \nu }\theta _{\kappa
\lambda },\, \\
\Pi _{\mu \nu ,\kappa \lambda }^{(L)} &=&\Lambda _{\mu \nu }\Lambda _{\kappa
\lambda },
\end{eqnarray}
\begin{eqnarray}
\Pi _{\mu \nu ,\kappa \lambda }^{(\omega L-a)} &=&\omega _{\mu \lambda
}\Lambda _{\nu \kappa }+\omega _{\nu \lambda }\Lambda _{\mu \kappa
},\,\,\,\,\Pi _{\mu \nu ,\kappa \lambda }^{(\omega L-b)}=\omega _{\mu \kappa
}\Lambda _{\nu \lambda }+\omega _{\nu \kappa }\Lambda _{\mu \lambda }, \\
\Pi _{\mu \nu ,\kappa \lambda }^{(\omega L)} &=&\omega _{\mu \nu }\Lambda
_{\kappa \lambda },\,\,\,\,\Pi _{\mu \nu ,\kappa \lambda }^{(L\omega
)}=\Lambda _{\mu \nu }\omega _{\kappa \lambda },
\end{eqnarray}
\begin{eqnarray}
\Pi _{\mu \nu ,\kappa \lambda }^{(\omega \Sigma -a)} &=&\omega _{\mu \nu
}\Sigma _{\kappa \lambda },\,\,\,\,\Pi _{\mu \nu ,\kappa \lambda }^{(\omega
\Sigma -b)}=\omega _{\mu \nu }\Sigma _{\lambda \kappa } ,\\
\Pi _{\mu \nu ,\kappa \lambda }^{(\Sigma \omega )} &=&\Sigma _{\mu \nu
}\omega _{\kappa \lambda }+\Sigma _{\nu \mu }\omega _{\kappa \lambda },
\end{eqnarray}
\begin{eqnarray}
\Pi _{\mu \nu ,\kappa \lambda }^{(L\Sigma -a)} &=&\Lambda _{\mu \nu }\Sigma
_{\kappa \lambda },\,\,\,\,\Pi _{\mu \nu ,\kappa \lambda }^{(L\Sigma
-b)}=\Lambda _{\mu \nu }\Sigma _{\lambda \kappa }, \\
\Pi _{\mu \nu ,\kappa \lambda }^{(\Sigma L)} &=&\Sigma _{\mu \nu }\Lambda
_{\kappa \lambda }+\Sigma _{\nu \mu }\Lambda _{\kappa \lambda },
\end{eqnarray}
The products between the usual spin operators for the subspace of symmetric rank-2 tensors are summarised as follows: 
\begin{eqnarray}
P^{i-a}P^{j-b} &=&\delta ^{ij}\delta ^{ab}P^{j-b},  \nonumber  \\
P^{i-ab}P^{j-cd} &=&\delta ^{ij}\delta ^{bc}P^{j-a},  \nonumber \\
P^{i-a}P^{j-bc} &=&\delta ^{ij}\delta ^{ab}P^{j-ac},   \label{17a}\\
P^{i-ab}P^{j-c} &=&\delta ^{ij}\delta ^{bc}P^{j-ac},  \nonumber
\end{eqnarray}
and they satisfy the following tensorial completeness relation: 
\begin{equation}
\left( P^{(2)}+P^{(1)}+P^{(0-s)}+P^{(0-w)}\right) _{\mu \nu ,\kappa \lambda
}=\frac{1}{2}\left( \eta _{\mu \kappa }\eta _{\nu \lambda }+\eta _{\mu
\lambda }\eta _{\nu \kappa }\right) .  \label{17b}
\end{equation}

After lengthy algebraic operations with the whole set of the operators presented above, we are ready to give the explicit form of $h_{\mu\nu}$-propagator in momentum space, which reads as below:
 
\begin{eqnarray}
\left\langle hh\right\rangle &=&\frac{i}{D}\left\{ 4D_{1}P^{(2)}+2\frac{N_{1}%
}{D_{1}\left( p^{2}-\widetilde\Lambda \alpha \right) }P^{(1)}-\frac{N_{2}}{%
D_{1}D_{2}}P^{(0-s)}+\frac{N_{3}D}{D_{2}\left( p^{2}-\widetilde\Lambda \alpha \right) }P^{(0-\omega )}\right.  \nonumber \\
&&+\sqrt{3}\frac{N_{4}}{D_{1}D_{2}}(P^{(0-s\omega )}+P^{(0-\omega
s)})+4\kappa ^{2}p^{2}S+12i\frac{\kappa ^{4}\lambda p^{4}}{D_{1}}\left( \Pi
^{\left( 1-a\right) }+\Pi ^{\left( 1-b\right) }\right)  \nonumber \\
&&+12\frac{\kappa ^{4}p^{6}}{D_{1}}\Pi ^{\left( 2\right) }-8\sqrt{3}i\frac{%
\kappa ^{4}\lambda p^{4}}{D_{1}}\left( \Pi ^{\left( \theta \Sigma -a\right)
}+\Pi ^{\left( \theta \Sigma -b\right) }+\Pi ^{\left( \Sigma \theta \right)
}\right)  \nonumber \\
&&-\left. 8\sqrt{3}\frac{\kappa ^{4}p^{6}}{D_{1}}\left( \Pi ^{\left( \theta
L\right) }+\Pi ^{\left( L\theta \right) }\right) \right\} , \label{propagator}
\end{eqnarray}
where $\lambda=v^\mu p_\mu$. We have suppressed the Minkowsky space indices from de  $h_{\mu \nu }$-field and from the spin-type operators. Moreover, we have defined: 
\begin{eqnarray}
D_{1} &=&\xi p^{4}+2p^{2}-2\widetilde\Lambda ; \\
D &=&\xi^{2}p^{8}+4\left( \xi +\kappa ^{4}v^{2}\right) p^{6}+4\left(1-\widetilde\Lambda \xi -
\kappa ^{4}\lambda ^{2}\right)p^{4}\nonumber\\
&&-8\widetilde\Lambda p^{2}+4\widetilde\Lambda ^{2}; \\
D_{2} &=&\left( \xi +3\sigma \right) p^{4}-p^{2}+\widetilde\Lambda;
\end{eqnarray}
\begin{eqnarray}
N_{1} &=&\alpha \xi ^{3}p^{12}+2\alpha \xi \left( 2\kappa ^{4}v^{2}+3\xi
\right) p^{10}+2\alpha \left( 4\kappa ^{4}v^{2}-2\kappa ^{4}\xi \lambda
^{2}-3\widetilde\Lambda \xi ^{2}+6\xi \right) p^{8} \nonumber\\
&&+2\left[ 4\alpha \left( 1-\kappa ^{4}\lambda ^{2}-\Lambda \kappa
^{6}v^{2}-3\widetilde\Lambda \xi \right) + 3\kappa ^{4}\lambda ^{2}\right]
p^{6}+2\alpha \widetilde\Lambda \left( 6\widetilde\Lambda \xi +\kappa
^{4}\lambda ^{2}-12\right) p^{4} \nonumber\\
&&+24\alpha \widetilde\Lambda ^{2}p^{2}-8\alpha \widetilde\Lambda ^{3};
\end{eqnarray}
\begin{eqnarray}
N_{2} &=&\xi ^{3}p^{12}+6\left( 2\kappa ^{4}\xi v^{2}+\xi ^{2}+4\kappa
^{4}v^{2}\sigma \right) p^{10}+6\left( 2\xi -4\kappa ^{4}\lambda ^{2}\sigma
-2\kappa ^{4}\xi \lambda ^{2}-\widetilde\Lambda \xi ^{2}\right) p^{8} \nonumber\\
&&+8\left[ 1-3\widetilde\Lambda \xi \right] p^{6}+12\widetilde\Lambda
\left( \Lambda \kappa ^{2}\xi -2\right) p^{4} \nonumber\\
&&+24\widetilde\Lambda ^{2}p^{2}-8\widetilde\Lambda ^{3};
\end{eqnarray}
\begin{equation}
N_{3}=4\alpha \left( \xi +3\sigma \right) p^{4}+\left( 3-4\alpha \right)
p^{2}+\alpha \widetilde\Lambda ; 
\end{equation}
\begin{eqnarray}
N_{4} &=&\xi ^{3}p^{12}+2\left( 2\kappa ^{4}\xi v^{2}+3\xi ^{2}\right)
p^{10}+2\left( 6\xi -12\kappa ^{4}\lambda ^{2}\sigma -6\kappa ^{4}\xi
\lambda ^{2}-3\Lambda \kappa ^{2}\xi ^{2}+4\kappa ^{4}v^{2}\right) p^{8} \nonumber\\
&&+8\left[ 1-\kappa ^{4}\widetilde\Lambda v^{2}-3\kappa ^{2}\Lambda \xi \right]
p^{6}+12\widetilde\Lambda \left( \widetilde\Lambda \xi -2\right) p^{4} \nonumber\\
&&+24\widetilde\Lambda ^{2}p^{2}-8\widetilde\Lambda ^{3}.
\end{eqnarray}
This is a very general expression with all parameters taken as non-trivial. We have to select only those cases for which ghosts and tachyons are absent. This imposes special contraints on the parameters as we are going to present in our final Section.
\section{Concluding Comments:}
With the general result above, we can focus on two specific cases.

By starting with a theory described by the Einstein-Hilbert plus
Chern-Simons terms, we get the following tree-level propagator:

\begin{eqnarray}
\left\langle hh\right\rangle  &=&\frac{i}{p^{2}\Delta}\left\{ 2P^{(2)}+2\left[ 
\frac{3}{4}\kappa ^{4}\lambda ^{2}+\alpha \Delta\right] P^{(1)}+P^{(0-s)}+\left(
4\alpha -3\right) \Delta P^{(0-\omega )}\right.   \nonumber \\
&&-\sqrt{3}\left( p^{2}\kappa ^{4}v^{2}+1\right) (P^{(0-s\omega
)}+P^{(0-\omega s)})+\kappa ^{2}S+\frac{3i}{2}\kappa ^{4}\lambda \left( \Pi ^{\left(
1-a\right) }+\Pi ^{\left( 1-b\right) }\right)   \nonumber \\
&&+\frac{3\kappa ^{4}p^{2}}{2}\Pi ^{\left( 2\right) }-\sqrt{3}i\kappa
^{4}\lambda \left( \Pi ^{\left( \theta \Sigma -a\right) }+\Pi ^{\left(
\theta \Sigma -b\right) }+\Pi ^{\left( \Sigma \theta \right) }\right)  
\nonumber \\
&&-\left. \sqrt{3}\kappa ^{4}p^{2}\left( \Pi ^{\left( \theta L\right) }+\Pi
^{\left( L\theta \right) }\right) \right\} ; \label{hh}
\end{eqnarray}
where $\Delta\left( p\right) =\kappa ^{4}v^{2}p^{2}-\kappa ^{4}\left( p\cdot
v\right) ^{2}+1.$ From (\ref{hh}), we notice that  $p^2=0$ is a pole and, so, a massless excitation is present despite the Lorentz-symmetry violating term.
It is now worthy to discuss the pole structure our propagator (\ref{hh}) exhibits. There is a factorisation of the zero mass pole, ($p^2=0$), while  non-trivial poles might appear as zeroes of the $\Delta(p)$-factor. In view of this result, the works of refs. \cite{Jackiw:2003pm, Jackiw:2003ri} should  be commented on, where Jackiw and Pi consider the particular case of a time-like vector, $v^\mu=(\frac{1}{\mu},\vec{0})$. They show that the violation of Lorentz symmetry, in such a case, is not felt, since its neat effect simply amounts to modifying the external sources in their spatial dependence. In our case, the factor $\Delta(p)$, also for the choice $v^\mu=(\frac{1}{\mu},\vec{0})$, looses its dependence on the $p^0$-component of the momentum: we have indeed $\Delta(p)=1-\frac{\kappa^4}{\mu^2}\vec{p}^2$; therefore, no other particle pole is present other than $p^2=0$. This then confirms the results by Jackiw and Pi. The $\Delta(p)$-factor, from which the energy dependence drops out in this case, may actually (as these authors propose) be absorbed into the redefinition of the external currents and only the massless $p^2=0$ - pole of the unmodified theory is actually present.\\

For $v^\mu$ space-like (let us take $v^\mu=(0;0,0,v)$), $\Delta(p)$ yields non-tachyonic massive poles, in agreement with the results found in \cite{Adam:2001ma, Adam:2002rg, BaetaScarpelli:2003yd} for the case of general Yang-Mills theories. Gravity may be formulated as a non-Abelian gauge theory, therefore, it is not surprising that, for $v^\mu$ space-like, massive poles may appear and they are of a non-tachyonic nature.

In theories with spontaneous gauge symmetry breaking in presence of a Lorentz symmetry breaking term, the competitive effect between the Higgs mechanism and the mass generation induced by the Lorentz-symmetry breaking term has been discussed \cite{Belich:2004pv}. In the particular case of gravitation, instead of an spontaneous gauge symmetry breaking, another way to generate mass to compete with the mass parameter of the Lorentz-symmetry breaking term would be through  the cosmological constant. This situation motivates a discussion of the excitation spectrum of the graviton propagator in presence of the cosmological constant and the background vector that violates the Lorentz symmetry.

By starting \ with a theory described by the Einstein-Hilbert,
cosmological constant and Chern-Simons terms, in (\ref{19}) we get:
\begin{eqnarray}
D_{1} &=&2p^{2}-2\Lambda \kappa ^{2}; \\
D &=&4p^4\Delta-8\Lambda \kappa ^{2}p^{2}+4\Lambda ^{2}\kappa ^{4}; \\
D_{2} &=&-p^{2}+\Lambda \kappa ^{2};
\end{eqnarray}
\begin{eqnarray}
N_{1} &=&8\alpha \kappa ^{4}v^{2}p^{8} \nonumber\\
&&+2\left[ 4\alpha \left( 1-\kappa ^{4}\lambda ^{2}-\Lambda \kappa
^{6}v^{2}\right) +3\kappa ^{4}\lambda ^{2}\right] p^{6}+2\alpha \kappa
^{2}\Lambda \left( \kappa ^{4}\lambda ^{2}-12\right) p^{4} \nonumber\\
&&+24\alpha \kappa ^{4}\Lambda ^{2}p^{2}-8\alpha \kappa ^{6}\Lambda ^{3};
\end{eqnarray}
\begin{equation} 
N_{2}=8p^{6}-24\kappa ^{2}\Lambda p^{4}+24\kappa ^{4}\Lambda
^{2}p^{2}-8\kappa ^{6}\Lambda ^{3};
\end{equation}
\begin{equation}
N_{3}=\left( 3-4\alpha \right) p^{2}+\alpha \kappa ^{2}\Lambda ;
\end{equation}
\begin{equation}
N_{4}=8\kappa ^{4}v^{2}p^{8}+8\left[ 1-\kappa ^{6}\Lambda v^{2}\right]
p^{6}-24\kappa ^{2}\Lambda p^{4}+24\kappa ^{4}\Lambda ^{2}p^{2}-8\kappa
^{6}\Lambda ^{3}.
\end{equation}

We conclude that there is no way that the cosmological constant and the violation of Lorentz symmetry may compete to cancel the effect of a massive graviton. On the other hand, in this case, the dispersion relation associated to the pole that corresponds to $D=0$ in eq.(\ref{propagator}) can be inspected to also yield an important information on the background vector, $v^\mu$. Actually, from

\begin{equation}
D\equiv 4\Delta\left[\left(p^2-\frac{\Lambda\kappa^2}{\Delta}\right)^2-\frac{\Lambda^2\kappa^4}{\Delta^2}(1-\Delta)\right]=0,
\end{equation}
and, whenever $\Delta\neq 0$,
\begin{equation}
\left(p^2-\frac{\Lambda\kappa^2}{\Delta}\right)^2=\frac{\Lambda^2\kappa^4}{\Delta^2}(1-\Delta).
\end{equation} 
Then $1-\Delta\geq0$; this condition reduces to $v^2p^2\leq (v.p)^2$, which is always satisfied whenever $v^\mu$ is space-like and $p^2=\mu^2>0$, which corresponds to the appearance of a massive time-like excitation. An issue which remains to be inspected is the inclusion of torsion in our discussion of the spectrum of Lorentz-symmetry violating gravity. 

A next step would consist in treating the vielbein and the spin connection (we are talking about the first-order formalism) as independent degrees of freedom to then reassess the whole set of gravity excitations that may show up in the spectrum. By virtue of the Lorentz-symmetry violation, we do not expect that the two approaches be equivalent and that this task is simply a matter of re-analysing the same results by means of another approach; this claim is based on the results of the work of Ref.\cite{Alexander:2008wi}. We expect that there may indeed appear an extra sector of (dynamical) massive gravitons accomodated in the spin connection sector. This problem is under consideration and we shall report on it in a near future \cite{Helayel}.   

\vspace{5mm}

\section*{Acknowledgements}
The authors express their gratitude to the Conselho Nacional
de Desenvolvimento Cient\'{i}fico e Tecnol\'{o}gico (CNPq-Brazil)
for the invaluable financial support.


\end{document}